\documentclass[showpacs,preprintnumbers,amsmath,amssymb,prb,twocolumn,superscriptaddress,floatfix]{revtex4}
\usepackage{graphicx}
\usepackage{bm}

\newcommand{\f}[1]{^{(#1)}} 
\newcommand{\ve}[1]{\mathbf{#1}}

\newcommand{\im}{{\rm i}}
\newcommand{\eq}[1]{Eq. (\ref{#1})}
\newcommand{\xy}{3D$xy$\ }

\newcommand{\beg}{\begin{eqnarray}}
\newcommand{\eee}{\end{eqnarray}}
\def\comment#1{}

\begin{document}

\title{Vortex sub-lattice melting in a two-component superconductor}

\author{E. Sm{\o}rgrav}
\affiliation{Department of Physics, Norwegian University of
Science and Technology, N-7491 Trondheim, Norway}

\author{J. Smiseth}
\affiliation{Department of Physics, Norwegian University of
Science and Technology, N-7491 Trondheim, Norway}

\author{E. Babaev}
\affiliation{Laboratory of Atomic and Solid State Physics, Cornell University, Ithaca, NY 14853-2501, USA}
\affiliation{Department of Physics, Norwegian University of Science and Technology, N-7491 Trondheim, Norway}

\author{A. Sudb{\o}}
\affiliation{Department of Physics, Norwegian University of
Science and Technology, N-7491 Trondheim, Norway}

\date{Received \today}

\begin{abstract}
We consider the vortex matter in a three-dimensional two-component superconductor 
with individually conserved condensates with different bare phase stiffnesses in 
a finite magnetic field, such as the projected superconducting state of liquid
metallic hydrogen. The ground state is a lattice of {\it composite}, 
i.e. co-centered, vortices in both order parameters. We investigate quantitatively 
two novel phase transitions when temperature is increased at fixed magnetic field. 
{\it i)} A ``vortex sub-lattice melting" phase transition where  vortices in the 
field with lowest phase stiffness (``light vortices") loose co-centricity with the  
vortices with large phase stiffness (``heavy vortices"), thus entering a liquid state.  
Remarkably, the structure factor of the light vortex sub-lattice vanishes 
{\it continuously}. This novel transition, which has no counterpart in one-component 
superconductors, is shown to be in the \xy universality class. Across this transition, 
the lattice of  heavy vortices {\it remains intact}.
{\it ii)} A first order melting transition of the  lattice of heavy vortices, with 
the novel feature that these are  {\it interacting with a background liquid of light 
vortices}. These findings are borne out in large-scale Monte Carlo simulations. 
\end{abstract}


\maketitle

Theories with multi-component bosonic scalar matter fields minimally coupled to a 
gauge field are of interest in a variety of condensed matter systems and beyond.
This includes  superconducting low temperature phases of 
light atoms  \cite{Ashcroft1999,egor2002,frac,smiseth2004,BSA,largepaper} under 
extreme enough pressures to produce liquid metallic states, easy-plane quantum 
antiferromagnets  \cite{senthil2003}, as well as other 
multiple-component superconductors \cite{egor2002,frac}. It also has applications 
in particle physics \cite{lee1973}. The projected liquid metallic states of hydrogen 
(LMH) \cite{Ashcroft1999} may soon be realized in high pressure experiments 
\cite{Datchi2000,Bonev}. Moreover, LMH is abundant in the interiors of the giant 
planets Jupiter and Saturn, where it is the origin of their magnetospheres 
\cite{Science}. At low temperatures, compressed liquid hydrogen is particularly 
interesting since it features prominent quantum fluctuations which lead to the 
possibility of a new state of matter, {a near ground state liquid metal} 
\cite{Ashcroft1999}. Its superconducting counterpart involves Cooper pairs of 
electrons and protons \cite{Ashcroft1999}, whence symmetry precludes Josephson 
coupling between  different condensate species. Resolving what happens to such 
a system {in a magnetic field} is now a matter of some urgency, due to new and 
detailed first principles calculations predicting LMH under extreme pressures 
of order $400GPa$ \cite{Bonev}. This is not far 
from experimentally achieved pressures of $320GPa$\cite{Datchi2000}, where 
hints of a maximum in the melting temperature versus pressure are evident. 
Magnetic-field experiments may very likely be exclusive probes 
to provide confirmation of LMH. A first study of the phase diagram of the projected LMH in 
magnetic fields has been presented, unveiling a phase diagram with rich structure \cite{BSA}. 
This raises issues of interest also in the broader domain of physics concerning the order 
and universality classes of possible phase transitions separating  phases of 
partially broken symmetries in quantum fluids. We report a study of this using a 
confluence of exact topological arguments and large-scale Monte-Carlo(MC) 
simulations. 
  
For general number of components $N$, the Ginzburg-Landau model is defined by the Lagrangian
\begin{eqnarray}
\label{gl_action}
{\cal{L}} = \sum_{\alpha=1}^N \frac{|\ve D \Psi_0^{(\alpha)}(\ve r)|^2}{2 M^{(\alpha)}}
+  V(\{\Psi_0^{(\alpha)}(\ve r)\}) + \frac{1}{2}(\nabla\times\ve{A}(\ve r))^2.
\end{eqnarray}
Here, $\{\Psi_0^{(\alpha)}(\ve r) \ | \ \alpha=1\dots N\}$ are complex scalar fields, 
$M^{(\alpha)}$ is the mass of the condensate species $\alpha$, and 
$\ve D = \nabla -\im e\ve{A}(\ve r)$. In LMH {\it each individual condensate is 
conserved}, consequently  $V(\{\Psi_0^{(\alpha)}(\ve r)\})$ is only a function of 
$|\Psi_0^{(\alpha)}(\ve r)|^2$. The model is studied in the phase-only approximation 
$\Psi_0^{(\alpha)}(\ve r) = |\Psi_0^{(\alpha)}|\exp [ \im\theta^{(\alpha)}(\ve r) ]$ 
where $|\Psi^{(\alpha)}_0|= {\rm const}$. Then, the $V$-term is a constant which may 
be omitted from the action \cite{BSA}.

For the discussions in this paper, another form of the action is useful. Introducing 
$|\psi^{(\alpha)}|^2 = |\Psi^{(\alpha)}_0|^2/M^{(\alpha)}$ and 
$\Psi^2 \equiv \sum_{\alpha=1}^N |\psi^{(\alpha)}|^2$,  \eq{gl_action} may be 
rewritten \cite{largepaper} in terms of {\it one} charged and $N-1$ neutral 
modes 
\begin{eqnarray}
{\cal L} & = & \frac{1}{2 \Psi^2} ~ \left( 
\sum_{\alpha = 1}^N |\psi^{(\alpha)}|^2 \nabla \theta^{(\alpha)}  - e \Psi^2 \ve A 
\right)^2  + \frac{1}{2}(\nabla\times\ve{A})^2 \nonumber \\
& + & \frac{1}{4 \Psi^2}  \sum_{\alpha,\beta=1}^N |\psi^{(\alpha)}|^2 ~ |\psi^{(\beta)}|^2 ~
 \left( \nabla (\theta^{(\alpha)}- \theta^{(\beta)}) \right)^2.
 \label{charge_neutral}
\end{eqnarray}
While \eq{gl_action} is convenient for MC simulations, \eq{charge_neutral} has advantages 
for analytical considerations, since the neutral and the charged modes are explicitly 
identified. 
Moreover, \eq{charge_neutral} is convenient for identifying various states of partially 
broken symmetry, emerging when an $N$-component system is subjected to an external magnetic 
field \cite{BSA,largepaper}.  

We now focus on the case $N=2$. In the case of LMH, $\Psi_0^{(1)}$ and $\Psi_0^{(2)}$ 
will denote protonic and electronic superconducting condensates, respectively, and 
hence 
$|\psi^{(1)}|^2 \ll |\psi^{(2)}|^2$.
In zero external magnetic field this system features a low-temperature phase-transition 
in the \xy universality class at $T_{\rm{c1}}$ where {superfluidity} is lost, followed 
at higher temperatures by an {inverted} \xy transition at $T_{\rm{c2}}$ where {superconductivity} 
and the Higgs mass of $\ve A$ (Meissner effect) is lost \cite{smiseth2004}. Here, we will 
consider the system in finite magnetic field at temperatures below $T_{\rm{c2}}$ \cite{BSA}.

We define a 
type-$\alpha$ vortex as a topological defect in $\Psi_0^{(\alpha)}$ associated with a 
non-trivial phase winding $\Delta \theta^{(\alpha)} = \pm 2 \pi$, whereas a 
{\it composite vortex} is a topological defect where type-$1$ and type-$2$ vortices 
coincide in space. 
\begin{figure}[htb]
\centerline{\scalebox{0.05}{\rotatebox{0.0}{\includegraphics{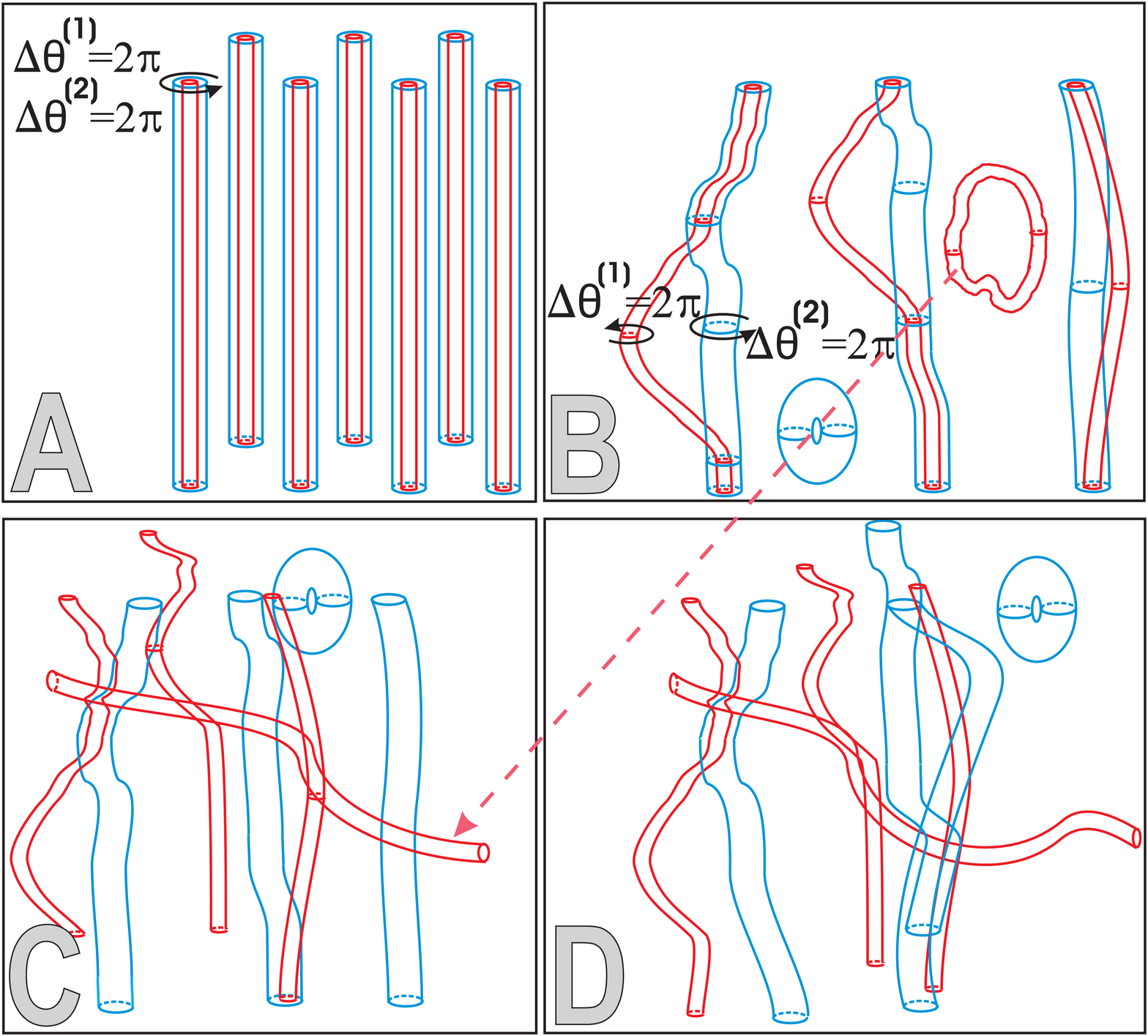}}}} 
\caption{ \label{phases} 
{\bf Panel A:} A type-II, $N=2$ superconductor at zero temperature in a magnetic field 
forms a lattice of composite vortices, i.e. co-centered type-$1$ (red) and 
type-$2$ (blue) vortices. 
{\bf Panel B:} Low-temperature  fluctuations in the composite  vortex lattice (VL). 
Thermal fluctuations generate closed loops of composite vortices and {\it local} 
splitting of field-induced composite vortices. This phase features superfluidity, 
as well as longitudinal superconductivity. {\bf Panel C:} A liquid of type-$1$ 
vortices  immersed in a background lattice of type-$2$ vortices.
There is a temperature region in low magnetic field when type-$1$ vortices form a vortex liquid 
and the corresponding vortex loops are proliferated, while type-$2$ vortices form a VL. 
Type-$1$ and type-$2$ vortices carry only a fraction of magnetic flux quantum in this 
state. Superfluidity is lost, longitudinal superconductivity is retained. The arrow from panel 
B to panel C illustrates  type-$1$ loop-proliferation. {\bf Panel D:} A vortex liquid of 
type-$1$ and type-$2$ vortices. Superfluidity and longitudinal superconductivity is lost, i.e. 
the normal phase\cite{BSA}. There is no type-$2$ loop proliferation 
going from panel C to panel D, since this is a first-order melting transition of the type-$2$ 
VL {\it in the background of a liquid of linetension-less type-$1$ vortices}.}
\end{figure}
At low temperatures the formation of a vortex lattice (VL) of {\it non-composite vortices} 
is forbidden because these vortices have a logarithmically divergent energy, whereas
{\it composite} vortices have finite energy \cite{frac,smiseth2004}. In a type-II 
$2$-component superconductor, therefore, a VL of { composite type-$1$ 
and type-$2$ vortices} is formed, illustrated in panel A of Fig. \ref{phases}. At 
elevated temperatures, the $2$-component system in a magnetic field will exhibit thermal 
excitations in the form of fractional-flux vortex loops similar to the case of zero magnetic 
field ${\bf B}={\bf{\nabla}} \times {\bf{A}} = 0$ \cite{frac,smiseth2004}. Since the 
field-induced  vortices are logarithmically bound states of constituent (elementary) 
vortices, the thermal fluctuations will induce a {\it local splitting} of composite 
vortices {in the form of two half-loops connected to a straight line} \cite{BSA}, as 
shown in panel B of Fig. \ref{phases}.

\begin{figure}[htb]
\centerline{\scalebox{0.13}{\rotatebox{0.0}{\includegraphics{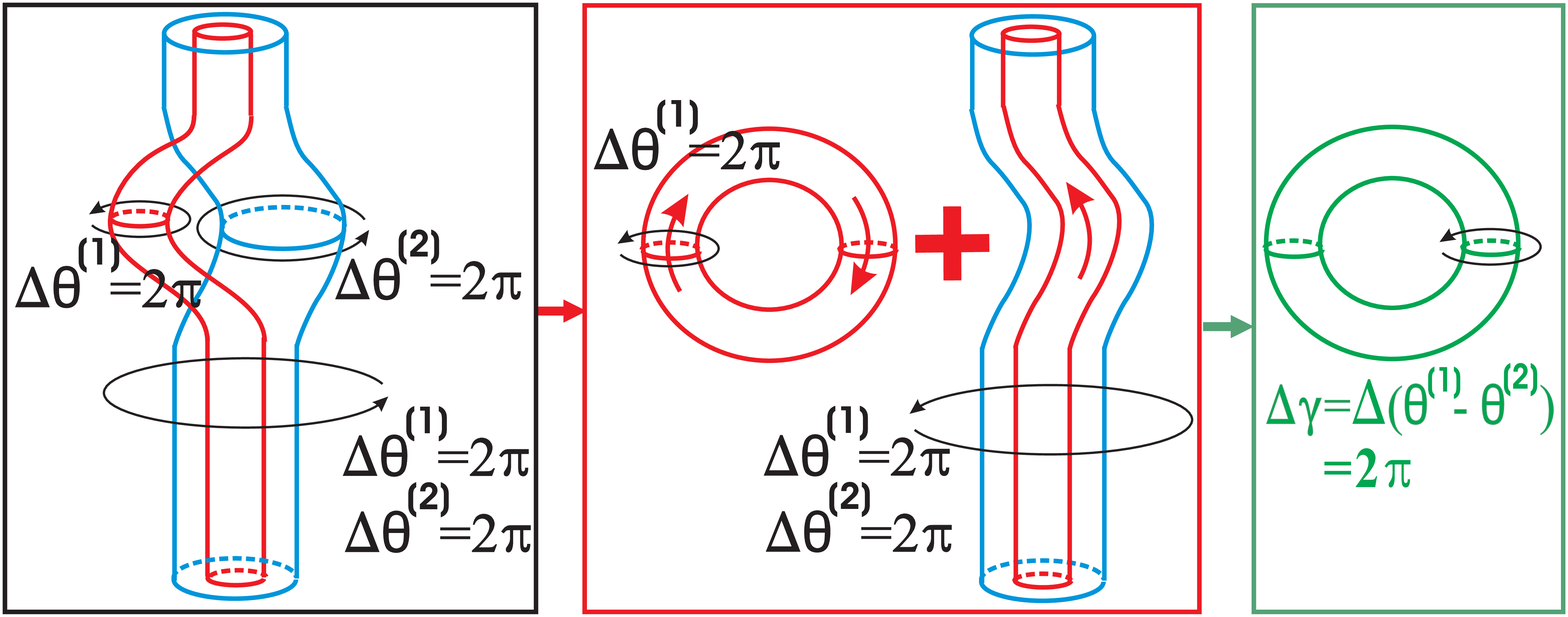}}}} 
\caption{ \label{splitting} 
Detailed illustration of the low-temperature thermal fluctuations in a VL of composite 
vortices. A local excursion of the vortex component with lowest bare phase stiffness 
(type-$1$ vortex) away from the composite VL may be viewed as a type-$1$ bound vortex 
loop superposed on the composite VL. 
The composite vortex line does not interact with a vortex with nontrivial winding in
$\Delta \gamma = \Delta(\theta^{(1)}-\theta^{(2)})$ \cite{largepaper}. A splitting 
transition of the composite VL, such as illustrated in going from panel B 
to panel C in Fig. \ref{phases}, may thus be viewed as a  {\it zero-field} vortex-loop 
proliferation of type-$1$ vortices, with a phase transition in the \xy universality 
class \cite{tesanovic1999,nguyen1999,Fossheim_Sudbo_book}.  }
\end{figure}

Consider now the processes illustrated in panel B of Fig. \ref{phases} for the case
$|\psi^{(1)}|^2 \ll |\psi^{(2)}|^2$, upon increasing the temperature beyond the 
low-temperature regime. We may view this process as a type-$1$ closed vortex loop 
superposed on a VL of (slightly) fluctuating composite vortices. 
An important point to notice is that a type-$\alpha$ vortex does not interact 
with a composite vortex by means of a neutral mode \cite{largepaper}. 
This follows from a topological argument that two split branches will feature 
nontrivialwinding in the composite neutral field $\theta^{(1)}-\theta^{(2)}$, 
while a composite vortex line does not. Hence, the splitting transition may be 
viewed as {\it a type-$1$ vortex loop-proliferation in a neutral superfluid}. 
This is illustrated in Fig. \ref{splitting}.
Thus, we may utilize the well-known results for the critical properties of the \xy model 
for neutral superfluids described as a vortex-loop proliferation 
\cite{tesanovic1999,nguyen1999,Fossheim_Sudbo_book}. 
This ``vortex sublattice melting" phase transition is therefore in the  \xy universality 
class \cite{tesanovic1999,nguyen1999,Fossheim_Sudbo_book}, not a first order melting 
transition. The resulting phase is one where superfluidity is lost and longitudinal 
superconductivity retained in the component $\Psi_0^{(2)}$, \cite{BSA} illustrated 
in panel C of Fig. \ref{phases}.

Apart from the sub-lattice melting transition, thermal fluctuations will produce a melting  
transition of the type-$2$ VL at a higher temperature. It is well known that 
sufficiently strong thermal fluctuations drive {\it a first-order} melting transition of the
Abrikosov lattice \cite{Fossheim_Sudbo_book} in $N=1$ superconductors. Due to the interplay with 
the proliferated type-$1$ vortices, a counterpart to this effect for the case $N=2$ when 
$|\psi^{(1)}|{}^2 \ne |\psi^{(2)}|{}^2$ is  more complex. The melting
temperature $T_{\rm M}(B)$ of the type-$2$ Abrikosov lattice  is suppressed with increasing 
magnetic field \cite{Fossheim_Sudbo_book}. At low enough magnetic fields, upon heating the 
system, the \xy type-$1$ vortex-loop transition at $T_{\rm{c1}}(B)$ will be encountered before
the melting transition of type-$2$ vortices at $T_{\rm M}(B)$. Above $T_{\rm M}(B)$, longitudinal 
superconductivity is also lost, whence we may infer that the vortex-liquid mixture of liberated 
type-$1$ and type-$2$ vortices is the normal metallic phase \cite{BSA}, depicted in panel D 
of Fig. \ref{phases}.

The above physical picture is borne out in large-scale MC simulations. 
We consider the model based on \eq{gl_action} for $N=2$ on an $L \times L \times L$ lattice 
with periodic boundary conditions for coupling constants $|\psi^{(1)}|^2=0.2$,
$|\psi^{(2)}|^2=2$, and $e^2=1/10$. The ratio  $|\psi^{(2)}|^2/|\psi^{(1)}|^2 = 10$ 
brings out one second order phase transition at $T_{c1}(B)$ in the \xy universality class well 
below the melting temperature $T_M(B)$ of the VL. In LMH 
$|\psi^{(2)}|^2/|\psi^{(1)}|^2 \approx 2000$, but the physical picture remains. For real 
estimates of $T_{c1}(B)$ and $T_M(B)$ in LMH, see Ref. \onlinecite{Ashcroft1999}. The 
Metropolis algorithm with local updating is used in combination with Ferrenberg-Swendsen
reweighting \cite{FSH}. System sizes as large as $L=96$ are used. The external magnetic 
field $\ve B$ studied is $B^x=B^y=0$, $B^z = 2\pi/32$, thus there are $32$ plaquettes 
in the  $(x,y)$-plane per flux-quantum. This is imposed by splitting the gauge field 
into a static part $\ve A_0$ and a fluctuating part $\ve A_{\rm fluct}$. The former 
is kept fixed to $(A_0^x,A_0^y(\ve r),A_0^z) = (0, 2\pi xf,0)$ where $f=1/32$ is the 
magnetic filling fraction, on top of which the latter field is free to fluctuate. 
Together with periodic boundary conditions on $\ve A_{\rm fluct}$,  the constraint 
$\oint_C(\ve A_{0}+\ve A_{\rm fluct}) d\ve l=2\pi fL^2$, where $C$ is a contour enclosing 
the system in the (x,y)-plane, is ensured. It is imperative to fluctuate $\ve A$, otherwise 
type-$1$ and type-$2$ vortices do not interact\cite{frac,smiseth2004,largepaper}. To investigate 
the transition at $T_{c1}$ we have performed finite size scaling (FSS) of the third moment 
of the action. The simulations are done by using vortices directly, similar to the procedure 
described in Ref. \onlinecite{smiseth2004}, but with a finite magnetic induction $B^z=2\pi/32$. 

\begin{figure}[htb]
\centerline{\scalebox{0.55}{\rotatebox{0.0}{\includegraphics{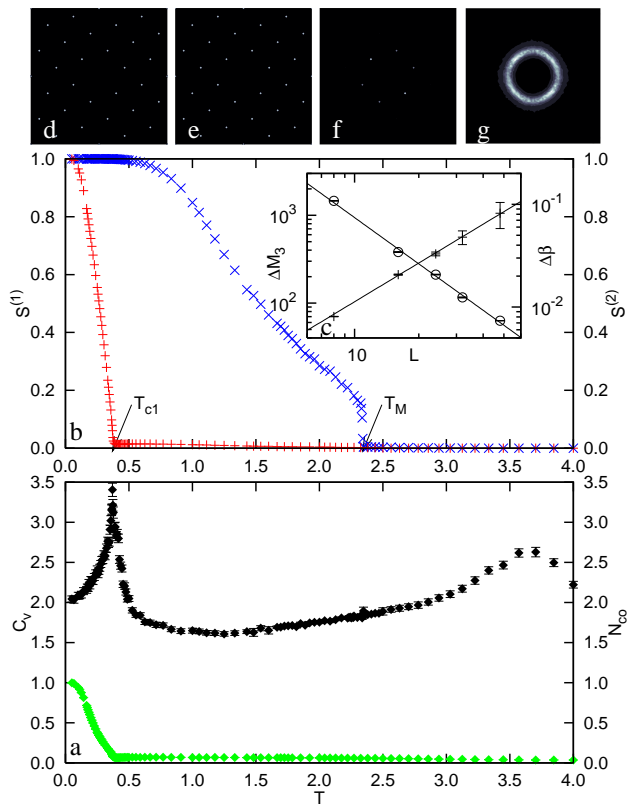}}}} 
\caption{ \label{MC_results} 
  MC results for $N=2$ $|\psi^{(1)}|^2=0.2$, $|\psi^{(2)}|^2=2$, and $e=1/\sqrt{10}$ 
  are presented. Panel a shows specific heat $C_v$ in black and co-centricity 
  $N_{\rm co}= N_{\rm co}^{+}-N_{\rm co}^{-}$ in green. Note how the specific 
  heat anomaly at $T_{c1}=0.37$, associated with the proliferation of
  type-$1$ vortices, matches the point at which $N_{\rm co}$ drops to
  zero, illustrating that type-$1$ vortices tear themselves off type-$2$
  vortices. The remnant of the zero-field anomaly in the specific heat can 
  be seen as a hump at $T \sim 3.6$. Panel b shows the structure factors  
  $S^{(1)}(\ve K)$ in red and $S^{(2)}(\ve K)$ in blue for
  the particular Bragg vector $\ve K =(\pi/4,-\pi/4)$. We see
  that $S^{(1)}(\ve K)$ vanishes continuously at $T_{\rm{c1}}$,
  while $S^{(2)}(\ve K)$ vanishes discontinuously at $T_{\rm M}=2.34$. 
  Panel c shows the FSS plots of the M3 from which the exponents 
  $\alpha=-0.02 \pm 0.05$ and $\nu=0.67 \pm 0.03$
  is extracted, showing that the sub-lattice melting is a critical phenomenon in the 
  \xy universality class. Panels d, e, f, and g are gray shade plots of the
  structurefactor $S^{(2)}(\ve k_\perp)$ for the temperatures $T_{\rm d}=0.35$,
  $T_{\rm e}=0.4$, $T_{\rm f}=1.66$, and $T_{\rm g}=2.85$,
  respectively. At $T_{\rm d}$ and
  $T_{\rm e}$ on both sides of the specific heat anomaly, the vortex lattice
  remains intact. The six fold Bragg symmetry is intact at $T_{\rm f}$,
  but is lost at $T_{\rm M}$ to give a ring pattern at $T_{\rm g}$, the
  hallmark of a vortex liquid.}  
\end{figure}

We compute the specific heat and the third moment of the action $S=\beta \int d\ve r\mathcal L$, 
defined as $M_3 = (\langle S^3 \rangle - \langle S \rangle^3 )/L^3$. Here, $\beta$ is inverse 
temperature. The peak to peak value of $M_3$ scales with system size as $L^{(1+\alpha)/\nu}$ and 
the width between the peaks scales as $L^{-1/\nu}$\cite{smiseth2002}. To probe the structural order 
of the vortex system we compute the planar structure function $S^{(\alpha)}(\ve k_\perp)$ 
of the {\it local vorticity} $\ve n^{(\alpha)}(\ve r) 
= \left( \ve \nabla \times  \left[ \ve \nabla \theta^{(\alpha)} -e ~  \ve  A  \right] \right)/2 \pi$, 
given by
\begin{equation}
S^{(\alpha)}(\ve k_\perp) = 
\frac{1}{(f L^3)^2}~ \langle | \sum_{\ve r} ~ n_z \f{\alpha}(\ve r) ~ e^{i \ve k_\perp \cdot \ve r_\perp} |^2 \rangle,
\end{equation}
where $\ve r$ runs over dual lattice sites and $\ve k_\perp$ is
perpendicular to $\ve B$. This function will exhibit sharp peaks for the characteristic 
Bragg vectors $\ve K$ of the type-$\alpha$ VL and will feature a ring-structure in its
corresponding liquid of type-$\alpha$ vortices. The signature of vortex sub-lattice melting will
be a transition from a six-fold symmetric Bragg-peak structure to a ring structure in $S^{(1)}(\ve K)$ 
while the peak structure remains intact in $S^{(2)}(\ve K)$. Furthermore, we compute the 
{\it vortex co-centricity} $N_{\rm{co}}$ of type-$1$ and type-$2$  vortices, defined as 
$N_{\rm co} \equiv  N^{+}_{\rm{co}} - N^{-}_{\rm{co}}$, where
\begin{equation}
N^{\pm}_{\rm{co}}
 \equiv \frac{\sum_{\ve r}|n_z\f 2(\ve r)|\delta_{n_z\f 1(\ve r),\pm
 n_z\f 2(\ve r)}}{\sum_{\ve r}|n_z\f 2(\ve r)|},
\end{equation}
where $\delta_{i,j}$ is the Kronecker-delta. The reason for considering $N_{\rm co}$ 
is that we then eliminate the effect of random overlap of vortices in the high-temperature 
phase $T > T_{\rm{c1}}$ due to vortex-loop proliferation, and focus on the {\it compositeness} 
of field-induced vortices. 
The quantity $N_{\rm{co}}$ is the fraction of type-$2$ vortex segments that are co-centered 
with type-$1$ vortices, providing a measure of the extent to which vortices of type-$1$ and 
type-$2$ form a {\it composite} vortex system. Hence, it probes the splitting processes 
visualized in panel B of Fig. \ref{phases} and in Fig. \ref{splitting}. The results are 
shown in Fig. \ref{MC_results}. 

The specific heat has a pronounced peak at $T_{\rm{c1}}$ associated with the \xy transition, 
and a broader less pronounced peak which is the finite field remnant of the zero-field 
inverted \xy transition \cite{nguyen1999}. Scaling of $M_3$ at $T_{\rm c1}$ shown in the inset 
c in Fig. \ref{MC_results} yields the critical exponents $\alpha =-0.02\pm0.05$ and 
$\nu=0.67\pm0.03$ in agreement with the \xy universality class. A novel result, not encountered 
in one-component superconductors, is that the structure function $S^{(1)}(\ve K)$ vanishes 
{\it continuously} as the temperature approaches $T_{\rm{c1}}$ from below, which is 
precisely the hallmark of the decomposition transition that separates the two types 
of vortex states depicted in panels B  and  C in Fig. \ref{phases}.
A related intriguing feature is the {\it vanishing} in the co-centricity $N_{\rm co}$ at 
$T_{\rm{c1}}$ as a function of temperature, which we will discuss in detail below. The 
first-order melting transition is seen to take place at  $T_{\rm M}$, where $S^{(2)}(\ve K)$ 
is seen to vanish discontinuously. This is the temperature at which the translational invariance 
is restored through melting of the type-$2$ VL. The resulting translationally invariant 
high-temperature phase is depicted in panel D of Fig. \ref{phases}. In the temperature interval 
$T < T_{\rm{c1}}$ the system {features superconductivity and superfluidity simultaneously} 
\cite{BSA}, since there is long-range order both in the charged and the neutral vortex modes. 
In the temperature interval $T_{\rm{c1}} < T < T_{\rm M}$ long-range order in the neutral mode 
is destroyed by loop-proliferation of type-$1$ vortices, hence { superfluidity} is lost 
\cite{BSA}. However, {longitudinal one-component superconductivity} is retained along the 
direction of the external magnetic field. For $T > T_{\rm M}$ superconductivity is also lost, 
hence this is the normal metallic state, which is a two-component vortex liquid. 

Next, we discuss these results in more detail. The most unusual and surprising feature is 
the continuous variation of $S^{(1)}(\ve K)$ with temperature, even at $T_{\rm{c1}}$ where 
it vanishes.
The explanation for this is the 
proliferation of type-$1$ vortices (which destroys the neutral superfluid mode) in the 
background of a composite VL, which the type-$1$ vortices essentially do not see, cf. Fig. 
\ref{splitting}. As far as the composite neutral Bose field $\theta^{(1)}-\theta^{(2)}$ 
is concerned, {\it it is precisely as if the composite VL were not present at all}. 
Hence, $S^{(1)}(\ve K)$ vanishes for a completely different reason than  $S^{(2)}(\ve K)$, 
namely due to {\it critical fluctuations, i.e. vortex-loop proliferation} in the condensate 
component with lowest bare stiffness. Such a phase transition does not completely restore 
broken translational invariance associated with a VL, since for the type-$2$ vortices 
{\it quite remarkably, the VL order survives the decomposition transition}, due to 
interaction between heavy vortices mediated by charged 
modes. The vanishing of $N_{\rm{co}}$ is particularly 
interesting, and finds a natural explanation within the framework of the above discussion. 
That is, for $T \ll T_{\rm{c1}}$, we have $N_{\rm{co}} \approx 1$, so the vortex system 
consists practically exclusively of composite vortices. As the temperature increases, thermal 
fluctuations induce excursions such as those illustrated in panel B of Fig. \ref{phases} and 
in Fig. \ref{splitting}, which reduces $N_{\rm co}^{\rm{+}}$ from its low-temperature value, 
reaching a {\it minimum} at $T_{\rm{c1}}$ and then {\it increase} for $T > T_{\rm{c1}}$. 
Conversely, $N_{\rm co}^{\rm{-}}$ remains essentially zero until  $T_{\rm{c1}}$, thereafter 
increasing monotonically. For temperatures above, but close to $T_{\rm{c1}}$, fluctuations 
in vortices originating in $\Delta \theta^{(2)}$ are still small, so the variations in 
$N_{\rm{co}}=N_{\rm co}^{\rm{+}}-N_{\rm co}^{\rm{-}}$ reflect thermal fluctuations 
in vortices originating in $\Delta \theta^{(1)}$. The increase of $N_{\rm co}^{\rm{\pm}}$ 
means that type-$1$ vortex loops  are thermally generated, and thus tend to {\it randomly} 
overlap more with the moderately fluctuating type-$2$ vortices. At their first order melting 
transition, type-$2$ vortices fluctuate only slightly. {\it Thus, the vanishing of $N_{\rm{co}}$ 
above $T_{\rm{c1}}$ reflects the increase in the density of thermally generated type-$1$ vortex 
loops in the background of a slightly fluctuating type-$2$ VL}, cf. panel C  in Fig. \ref{phases}.

In summary, we have investigated the orders and universality classes of thermally driven phase 
transitions in a two-component vortex system in a magnetic field. We find two phase transitions 
for the fields we consider in this paper (low-field regime). {\it i)} A \xy phase-transition 
associated with melting of the VL 
originating in the phases with lowest stiffness, which may be viewed as vortex-loop proliferation 
taking place in the background of a composite VL. The structure function associated with the 
type-$1$ vortices  vanishes {\it continuously} at the transition. This phase transition 
has no counterpart in a one-component superconductor. {\it ii)} A first-order VL 
melting associated with restoration of translational invariance of the type-$2$ vortex system. 
The corresponding structure function  vanishes {\it discontinuously} at the transition, which 
however takes place in the background of a {\it liquid of linetension-less type-$1$ vortices.}  
This also sets the type-$2$ VL melting apart from the corresponding phenomenon in one-component 
superconductors.

This work was funded by the 
Norwegian High Performance Computing Program, by the Research Council of Norway, Grant
Nos. 157798/432, 158518/431, and 158547/431 (NANOMAT), by STINT and the Swedish Research 
Council, and by the US National Science Foundation DMR-0302347. We acknowledge 
collaborations with N. W. Ashcroft (Ref. \onlinecite{BSA}).

\end{document}